# SECURED ONTOLOGY MAPPING

Manjula Shenoy.K[1], Dr.K.C.Shet[2], Dr. U.Dinesh Acharya[1]

[1]Department of Computer Engineering, Manipal University, MIT, Manipal
`manju.shenoy@manipal.edu, dinesh.acharya@manipal.edu`
[2]Department of Computer Engineering, NITK, Suratkal
`kcshet@rediffmail.com`

*ABSTRACT*

*Today's market evolution and high volatility of business requirements put an increasing emphasis on the ability for systems to accommodate the changes required by new organizational needs while maintaining security objectives satisfiability. This is all the more true in case of collaboration and interoperability between different organizations and thus between their information systems. Ontology mapping has been used for interoperability and several mapping systems have evolved to support the same. Usual solutions do not take care of security. That is almost all systems do a mapping of ontologies which are unsecured. We have developed a system for mapping secured ontologies using graph similarity concept. Here we give no importance to the strings that describe ontology concepts ,properties etc. Because these strings may be encrypted in the secured ontology. Instead we use the pure graphical structure to determine mapping between various concepts of given two secured ontologies. The paper also gives the measure of accuracy of experiment in a tabular form in terms of precision, recall and F-measure.*

## 1. INTRODUCTION

Researchers have developed several tools that enable organizations to share information, largely, most of these have not taken into the account the necessity of maintaining privacy and confidentiality of data and metadata of the organizations who want to share information. Consider the scenario of two different country military wanting to share information about a mission at hand while preserving the privacy of their systems. To the best of our knowledge current systems do not allow this type of information sharing.

Need for secured information sharing also exists for intra organizational information sharing too. Within the organizations different departments may use different systems which are autonomously constructed . The secure interoperability may be required here too.

Privacy should be maintained for both data and metadata. Metadata describes how data is organized (data schema) , how access are controlled in the organization( the internal access control policy and role hierarchies) and the semantics of the data used in the organization(ontology).

Organizations looking to interoperate are largely using metadata like ontologies to capture the semantics of the terms used in the information sources maintained by the organizations. Normally it has been assumed that these ontologies will be published by the organizations. Published ontologies from different organizations are mapped and matching rules are generated. Queries to information sources are rewritten using these matching rules so that vocabulary used in the query matches with the vocabulary of information source.

Unlike the traditional way some organizations may not like to publish their metadata or share it with other external users. Yet they want interoperation. In this case the privacy of the metadata must be preserved. The external user should not have access to ontologies in cleartext. So ontologies may be encrypted and then published. The mapping system should now be able to recognize mapping in this encrypted ontology. Here we present one such system.

## 2. RELATED WORK

The present ontology mapping systems can be classified into the following categories.

1. Word Similarity based: Here matching is performed based on similarity of words describing concepts, properties or names of concepts and properties occurring in the ontology.[4]
2. Structure based: Here structure of ontologies has been used for matching concepts.[5][6][7].
3. Instance based: These take the instances under concepts to find matching.[8]. These methods are further subdivided into Opaque and pattern based. In Opaque instance matching we use statistical properties like distribution ,entropy and mutual information etc. In Pattern based method instance pattern are matched.
4. Inference Based: The semantics of concepts under ontologies are expressed as rules in a logical language and then the matching is performed using an inference engine.
There are also hybrid algorithms for matching ontologies.
[1] discusses need for secured data sharing in or among organization and [2] explains need for secured data mining. [3] proposes two methods for privacy preserving ontology matching. One of which is semi-automatic. And the other requires the dictionaries or thesauri or corpuses to be encrypted. Our method falls purely under structure based ontology matching which can be applied to encrypted ontologies. [4] defines a graph matching technique we used, in the literature.

## 3. GRAPH MATCHING TECHNIQUE USED
### 3.1. Generalizing hubs and authorities[17]

Efficient web search engines such as Google are often based on the idea of characterizing the most important vertices in a graph representing the connections or links between pages on the web. One such method, proposed by Kleinberg [16], identifies in a set of pages relevant to a query search the subset of pages that are good *hubs* or the subset of pages that are good *authorities*. For example, for the query "university," the home-pages of Oxford, Harvard, and other universities are good authorities, whereas web-pages that point to these home-pages are good hubs. Good hubs are pages that point to good authorities, and good authorities are pages that are pointed to by good hubs. From these implicit relations, Kleinberg derives an iterative method that assigns an "authority score" and a "hub score" to every vertex of a given graph. These scores can be obtained as the limit of a converging iterative process, which is described in section below. Let $G = (V,E)$ be a graph with vertex set $V$ and with edge set $E$ and let $h_j$ and $a_j$ be the hub and authority scores of vertex $j$. We let these scores be initialized by some positive values and then update them simultaneously for all vertices according to the following *mutually reinforcing relation*: the hub score of vertex $j$ is set equal to the sum of the authority scores of all vertices pointed to by $j$, and, similarly, the authority score of vertex $j$ is set equal to the sum of the hub scores of all vertices pointing to $j$:

$$\begin{cases} h_j & \leftarrow & \sum_{i:(j,i)\in E} a_i, \\ a_j & \leftarrow & \sum_{i:(i,j)\in E} h_i. \end{cases}$$

Let *B* be the matrix whose entry (*i*, *j*) is equal to the number of edges between the vertices *i* and *j* in *G* (the *adjacency matrix* of *G*), and let *h* and *a* be the vectors of hub and authority scores. The above updating equations then take the simple form

$$\begin{bmatrix} h \\ a \end{bmatrix}_{k+1} = \begin{bmatrix} 0 & B \\ B^T & 0 \end{bmatrix} \begin{bmatrix} h \\ a \end{bmatrix}_k, \qquad k = 0, 1, \ldots,$$

which we denote in compact form by

$$x_{k+1} = M\, x_k, \qquad k = 0, 1, \ldots,$$

Where

$$x_k = \begin{bmatrix} h \\ a \end{bmatrix}_k, \qquad M = \begin{bmatrix} 0 & B \\ B^T & 0 \end{bmatrix}.$$

Notice that the matrix *M* is symmetric and nonnegative. We are interested only in the relative scores and we will therefore consider the *normalized* vector sequence

$$z_0 = x_0 > 0, \quad z_{k+1} = \frac{M z_k}{\|M z_k\|_2}, \qquad k = 0, 1, \ldots,$$

Where $\|..\|_2$ is the Euclidean vector norm. Notice that the above matrix *M* has the property that

$$M^2 = \begin{bmatrix} BB^T & 0 \\ 0 & B^T B \end{bmatrix},$$

and from this equality it follows that, if the dominant invariant subspaces associatedwith $BB^T$ and $B^T B$ have dimension 1, then the normalized hub and authority scores are simply given by the normalized dominant eigenvectors of $BB^T$ and $B^T B$. This is the definition used in [16] for the authority and hub scores of the vertices of *G*. The arbitrary choice of $z0 = \mathbf{1}$ made in [16] is shown here to have an extrenal norm justification. Notice that when the invariant subspace has dimension 1, then there is nothing particular about the starting vector **1**, since any other positive vector *z*0 would give the same result. We now generalize this construction. The authority score of vertex *j* of *G* can be thought of as a similarity score between vertex *j* of *G* and vertex *authority* of the graph
    hub➔authority

and, similarly, the hub score of vertex *j* of *G* can be seen as a similarity score between vertex *j* and vertex *hub*. The mutually reinforcing updating iteration used above can be generalized to graphs that are different from the hub–authority structure graph.

The idea of this generalization is easier to grasp with an example; we illustrate it first on the path graph with three vertices and then provide a definition for arbitrary graphs. Let *G* be a graph with edge set *E* and adjacency matrix *B* and consider the *structure graph*
        1➔2➔3

With each vertex *j* of *G* we now associate three scores *xi*1, *xi*2, and *xi*3, one for each vertex of the structure graph. We initialize these scores with some positive value and then update them according to the following mutually reinforcing relation:

$$\begin{cases} x_{i1} & \leftarrow & \sum_{j:(i,j)\in E} x_{i2}, \\ x_{i2} & \leftarrow & \sum_{j:(j,i)\in E} x_{i1} & + \sum_{j:(i,j)\in E} x_{i3}, \\ x_{i3} & \leftarrow & \sum_{j:(j,i)\in E} x_{i2}, \end{cases}$$

or, in matrix form (we denote by **x**j the column vector with entries *xij* ),

$$\begin{bmatrix} \mathbf{x}_1 \\ \mathbf{x}_2 \\ \mathbf{x}_3 \end{bmatrix}_{k+1} = \begin{bmatrix} 0 & B & 0 \\ B^T & 0 & B \\ 0 & B^T & 0 \end{bmatrix} \begin{bmatrix} \mathbf{x}_1 \\ \mathbf{x}_2 \\ \mathbf{x}_3 \end{bmatrix}_k, \quad k = 0, 1, \ldots,$$

which we again denote *xk*+1 = *Mxk*. The situation is now identical to that of the previous example and all convergence arguments given there apply here as well. We now come to a description of the general case. Assume that we have two directed graphs *GA* and *GB* with *nA* and *nB* vertices and edge sets *EA* and *EB*. We think of *GA* as a structure graph that plays the role of the graphs *hub − authority* and 1 − 2 − 3 in the above examples. We consider real scores *xij* for *i* = 1, . . . , *nB* and *j* = 1, . . . , *nA* and simultaneously update all scores according to the following updating equations:

$$x_{ij} \leftarrow \sum_{r:(r,i)\in E_B,\ s:(s,j)\in E_A} x_{rs} + \sum_{r:(i,r)\in E_B,\ s:(j,s)\in E_A} x_{rs}.$$

This equation can be given an interpretation in terms of the product graph of *GA* and *GB*. The *product graph* of *GA* and *GB* is a graph that has *nA.nB* vertices and that has an edge between vertices (*i*1, *j*1) and (*i*2, *j*2) if there is an edge between *i*1 and *i*2 in *GA* and there is an edge between *j*1 and *j*2 in *GB*. The above updating equation is then equivalent to replacing the scores of all vertices of the product graph by the sum of the scores of the vertices linked by an outgoing or incoming edge. Equation can also be written in more compact matrix form. Let *Xk* be the *nB* × *nA* matrix of entries *xij* at iteration *k*. Then the updating equations take the simple form

$$X_{k+1} = BX_k A^T + B^T X_k A, \quad k = 0, 1, \ldots,$$

where *A* and *B* are the adjacency matrices of *GA* and *GB*. This equation is further revised by Laure Ninove [18] as follows Where $X_K$ is replaced by $S_k$

$$\frac{BS_k A^t + B^t S_k A}{\| BS_k A^t + B^t S_k A \|}$$

## 4.SECURED ONTOLOGY MATCHING.
### 4.1. Graph Similarity Measure Matrix

First we explain the graph matching technique we used. Consider the two graphs Ga and Gb shown in Figure 1. Suppose we want to match vertex 1 of Ga with vertex 4 of Gb , we need to find how much similar the vertices 2 of Ga and 2 of Gb , and 2 of Ga and 1 of Gb.

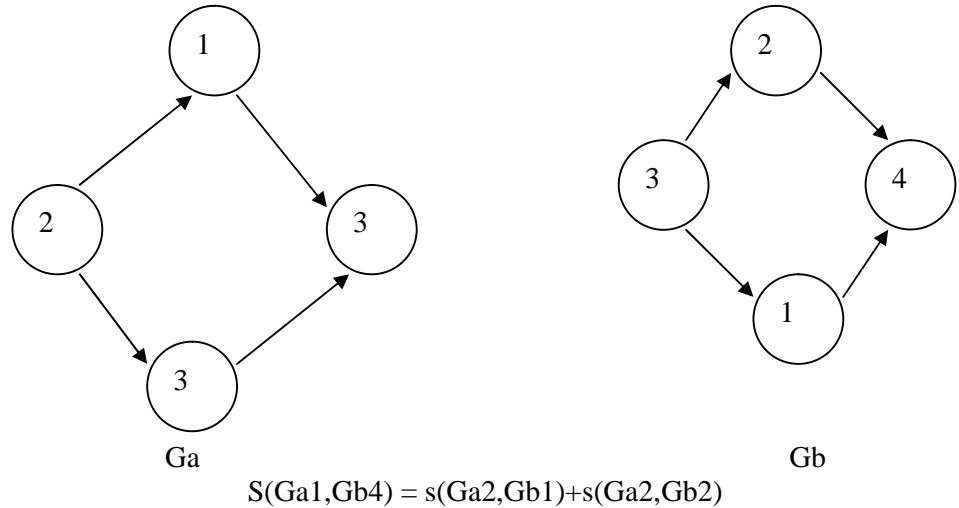

S(Ga1,Gb4) = s(Ga2,Gb1)+s(Ga2,Gb2)

Figure 1. Graphs to be matched

If A is the adjacency matrix of Ga and B is the adjacency matrix of Gb and S is the similarity matrix defined as follows between vertices we can get the total similarity matrix between individual vertices can be calculated using the formula

$$\frac{BSA^t + B^tSA}{\| BSA^t + B^tSA \|}$$

Here $A^t$ stands for transpose of A. S is the initial similarity matrix. The size of S is nXm.
Where m is number of concepts in first ontology and n is number of ontology concepts in second. The secured mapping method generates adjacency matrices based on hierarchical relationship of concepts of the encrypted ontologies as per the following algorithm.

---

Algorithm 1. Generating Adjacency matrix for the encrypted ontology given
Let O be the ontology given and A for adjacency matrix. If n is the number of concepts in ontology O then A has order nXn.
1. Initialize A [i][j]=0 for all i and j between 0 and n.
2. For i= 1to n
    Begin
        Str=get $i^{th}$ concept of O
        Collection = get all super classes of Str.
        For each Object x in the Collection
            Begin
                For j = 1 to n
                If jth concept of O matches with x then A[i][j]=1;
            End
        End

S is the unity matrix initially.

## 4.2. DegreeDifference Similarity(DDS) Matrix

The degree of a node is the number of edges connected to this node. In the algorithm we first compute SSNdegree(sum of self and neighbor degree) of every node. This is the sum of the nodes degree plus its neighbors degree. In figure 1 For Ga, SSNdegree for node 1 is 2+2+2=6. For the nodes to be mapped we find difference between SSNdegree and subtract it from maximum degree of the graph and call it Degree Difference Similarity value. For whole matching problem these values become part of the similarity matrix.

## 4.3. NodeAttributeSimilarity (NAS)Matrix

This indicates how many neighbours of node 1 match neighbors of node 2. For each neighbor of node1, we examine which neighbor of node 2 can be mapped based on same attribute name. If such a mapping is present NAS (initialized to 0) is incremented by 1. For every node pairs we compute NAS and express it as a matrix for whole matching problem.

## 4.4. Edge AttributeSimilarity(EAS)Matrix

This is similar to NAS. Rather than comparing nodes we compare edges here. i.e. For each edge of node1 we try to map a corresponding edge of node 2 which has the same attribute. When EAS is computed for every pair of nodes in two graphs we get a matrix.

## 4.5. Bayesian Belief Network

Here We describe an approach for ontology matching using Bayesian network. Our approach described here does not use Bayesian network to detect mappings. Instead we apply network to learn relationship between different similarity measures treated as different mapping methods and then to choose the best mapping. BBN's considered are assumed to contain nodes one per similarity measure and one output node representing the final output. This will allow us not only to combine the methods (in the probabilistic framework) but also to talk about conditionally independent methods, a minimal required subset of methods and the like. The input to the process of BBN training for ontology mapping are positive and negative examples with results of individual methods. The positive examples correspond to pairs for which mapping has previously been established, while the negative ones are (all or a subset of) pairs that have been identified as non-matching. Then CPTs and the structure are learnt using famous K2 algorithm. In the phase of using the trained BBN, the mapping justifications for unseen cases (pairs of concepts) are counted and inserted into the BBN as evidence. The result of alignment is calculated via propagation of this evidence.

## 4.6. K2 Algorithm[19]

This is to find the most probable Bayes-network structure given a database.
D – a database of cases   Z – the set of variables represented by D , $B_{si}$ , $B_{sj}$ – two bayes network structures containing exactly those variables that are in Z.  By computing such ratios for pairs of bayes network structures, we can rank order a set of structures by their posterior probabilities. Based on four assumptions, the paper introduces an efficient formula for computing $P(B_s,D)$, let B represent an arbitrary bayes network structure containing just the variables in D.
Assumption 1: The database variables, which we denote as Z, are discrete
Assumption 2: Cases occur independently, given a bayes network model

Assumption 3: There are no cases that have variables with missing values
Assumption 4: The density function $f(B_p|B_s)$ is uniform. $B_p$ is a vector whose values denotes the conditional-probability assignment associated with structure $B_s$
D - dataset, it has m cases(records)
Z - a set of n discrete variables: $(x_1, \ldots, x_n)$
$r_i$ - a variable $x_i$ in Z has $r_i$ possible value assignment: $v_{i1}..v_{ir_i}$

$B_s$ - a bayes network structure containing just the variables in Z
$\pi_i$ - each variable $x_i$ in $B_s$ has a set of parents which we represent with a list of variables $\pi_i$
$q_i$ - there are has unique instantiations of $\pi_i$
$w_{ij}$ - denote jth unique instantiation of $\pi_i$ relative to D.
$N_{ijk}$ - the number of cases in D in which variable $x_i$ has the value of
and $\pi_i$ is instantiated as $w_{ij}$.

$$N_{ij} = \sum_{k=1}^{r_i} N_{ijk}$$

$$\frac{P(B_{S_i} | D)}{P(B_{Sj} | D)} = \frac{\frac{P(B_{S_i}, D)}{P(D)}}{\frac{P(B_{Sj}, D)}{P(D)}} = \frac{P(B_{Si}, D)}{P(B_{Sj}, D)}$$

$$P(B_s, D) = P(B_s) \prod_{i=1}^{n} \prod_{j=1}^{q_i} \frac{(r_i - 1)!}{(N_{ij} + r_i - 1)!} N_{ij} \prod_{k=1}^{r_i} N_{ijk}!$$

Three more assumptions to decrease the computational complexity to polynomial-time:
<1> There is an ordering on the nodes such that if $x_i$ precedes $x_j$, then we do not allow structures in which there is an arc from $x_j$ to $x_i$.
<2> There exists a sufficiently tight limit on the number of parents of any nodes
<3> $P(\pi_i \to x_i)$ and $P(\pi_j \to x_j)$ are independent when $i \neq j$.

$$\max_{B_s}[P(B_s, D)] = \prod_{i=1}^{n} [P(\pi_i \to x_i) \prod_{j=1}^{q_i} \frac{(r_i - 1)!}{(N_{ij} + r_i - 1)!} N_{ij} \prod_{k=1}^{r_i} N_{ijk}!$$

Use the following functions:

$$g(i, \pi_i) = \prod_{j=1}^{q_i} \frac{(r_i - 1)!}{(N_{ij} + r_i - 1)!} \prod_{k=1}^{r_i} N_{ijk}!$$

Where the $N_{ijk}$ are relative to $\pi_i$ being the parents of $x_i$ and relative to a database D
Pred($x_i$) = $\{x_1, \ldots x_{i-1}\}$
It returns the set of nodes that precede $x_i$ in the node ordering.

## 5. RESULTS

The evaluation of the proposed system above is carried out for OAEI systematic benchmark suite. Since we compare for equality of names, and give importance to structure we need not encrypt the ontology for study of evaluation measures. The evaluation measures we considered are Precision, Recall and F-measure. Precision gives the ratio of correctly found correspondences over the total number of returned correspondences. If R is the reference alignment and A is the found alignment then the ratio for precision is

$$P(A, R) = \frac{|R \cap A|}{|A|}.$$

Recall is the ratio of correctly found correspondences to the total number of expected correspondences. The formula is

$$R(A, R) = \frac{|R \cap A|}{|R|}.$$

The following formula is used for finding F-measure.

$$M_\alpha(A, R) = \frac{P(A, R) \times R(A, R)}{(1 - \alpha) \times P(A, R) + \alpha \times R(A, R)}.$$

Here   is between 0 and 1. If   is 1 F-measure is same as precision otherwise if it is 0 then F-measure is same as recall. Usually it is taken as 0.5.

Table 1 gives the dataset and the results of experiments in terms of evaluation measures stated above.

Table 1

| Benchmark test no | Precision | Recall | F-measure |
| --- | --- | --- | --- |
| 1xx | 1 | 0.9 | 0.95 |
| 2xx | 1 | 0.89 | 0.94 |
| 3xx | 0.9 | 0.87 | 0.88 |

## 6. CONCLUSIONS

Maintaining privacy in interoperation systems is becoming increasingly important. Ontology matching is the primary means of resolving semantic heterogeneity. Ontology matching helps establish semantic correspondence rules that are used for query rewriting and translation in interoperation systems. For information systems that want maximum privacy, the privacy of their ontologies must be maintained. Our system gives a method to map ontologies which are secured.

## Authors

Manjula Shenoy.K is currently working as a    Assistant professor (Sel.Grade) at CSE Department, Manipal Institute of Technology, Manipal   University,Manipal.

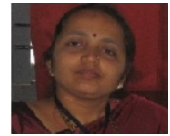

Dr. K.C. Shet is a Professor in Department of Computer Engineering, National Institute of  Technology ,Suratkal. He has several Journal and Conference papers to his credit.

Dr. U.Dinesh Acharya is Professor and Head of the department of Computer Science and Engineering, Manipal Institute of Technology , Manipal University,Manipal. He has several Journal and Conference papers to his credit.

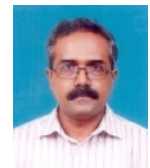